\documentclass[journal,10pt]{IEEEtran}
\usepackage{multirow}
\usepackage{tabu}
\usepackage{makecell}
\usepackage{color}
\usepackage[table]{xcolor}
\usepackage{graphicx}
\usepackage{bm}
\usepackage{balance}
\usepackage{amsmath,amssymb,amsfonts}

\usepackage{bbding}

\begin{document}

\title{Anomaly Detection in Intra-Vehicle Networks}


\author{\IEEEauthorblockN{Ajeet Kumar Dwivedi}\\
\IEEEauthorblockA{
Western University, London, Ontario, Canada \\
e-mail: adwived3@uwo.ca}

}

\markboth{}
{}

\maketitle

\begin{abstract}
The progression of innovation and technology and ease of inter-connectivity among networks has allowed us to evolve towards one of the promising areas, the Internet of Vehicles. Nowadays, modern vehicles are connected to a range of networks, including intra-vehicle networks and external networks. However, a primary challenge in the automotive industry is to make the vehicle safe and reliable; particularly with the loopholes in the existing traditional protocols, cyber-attacks on the vehicle network are rising drastically. Practically every vehicle uses the universal Controller Area Network (CAN) bus protocol for the communication between electronic control units to transmit key vehicle functionality and messages related to driver safety. The CAN bus system, although its critical significance, lacks the key feature of any protocol authentication and authorization. Resulting in compromises of CAN bus security leads to serious issues to both car and driver safety. This paper discusses the security issues of the CAN bus protocol and proposes an Intrusion Detection System (IDS) that detects known attacks on in-vehicle networks. Multiple Artificial Intelligence (AI) algorithms are employed to provide recognition of known potential cyber-attacks based on messages, timestamps, and data packets traveling through the CAN. The main objective of this paper is to accurately detect cyberattacks by considering time-series features and attack frequency. The majority of the evaluated AI algorithms, when considering attack frequency, correctly identify known attacks with remarkable accuracy of more than 99\%. However, these models achieve approximately 92\% to 97\% accuracy when timestamps are not taken into account. Long Short Term Memory (LSTM),  Xgboost, and SVC have proved to the well-performing classifiers.
\end{abstract}


\IEEEpeerreviewmaketitle

\section{Introduction}
The modern world is becoming increasingly reliant on technologies such as the Internet of Vehicles (IoV), connected vehicles (CV), and autonomous vehicles (AV). IoV is the principal vehicular communication framework that allows vehicles to communicate effectively with other IoV entities such as infrastructures, pedestrians, and smart devices [1][2]. Vehicle-To-Everything (V2X) innovation permits cutting-edge vehicles to communicate with other vehicles, road-side foundations, and street users [2][3]. Sensor readings are used by fully automated vehicles to make short- and long-term driving decisions. The hi-tech architecture of these vehicles improves communication between the sensors. Whether it is a self-driving car or an autonomous vehicle, these emerging technologies not only enhance efficiency but also provide a safe mode of transportation. Intra-vehicle networks (IVN) and external vehicular networks make up the majority of IoV [3]. 

Numerous electronic control units (ECUs) are used in IVNs to actualize different functionalities [3]. Modern cars comprise around 30 to 100 ECUs. These ECUs receive unique CAN ID messages during communications, and they are responsible for integrating, operating, and analyzing loads of internal vehicle components and electrical systems. The controller area network (CAN), a standard bus protocol for in-vehicle connectivity, connects all ECUs in a vehicle to transfer messages and perform actions [4]. Each component has the ability to share data with the others. For example, an Anti-Lock Braking System (ABS) in a car displays information about the brakes and airbags. Essentially, all of the components are linked and can communicate with one another. Telematics and infotainment systems are two significant sections of the communication of ECUs [5]\cite{tree}. Telematics equipment such as On-Board Diagnostics (OBD), black boxes, and other telematics devices are part of the telematics system. These are tracking devices that are installed in vehicles that allow for the broadcast and storage of real-time telemetry insights data via wireless connections. Also, vehicles' inbuilt modem and troubleshooting tools store the data for insights \cite{tree}[7]. All of a car's communication and entertainment functions are controlled by an infotainment system [7]. The infotainment system, which is placed in the car in the form of Touchscreens, provides interfaces for information and enjoyment, whether it is phone calls or music advice [8]\cite{mth}. The best part is that they let drivers access their phone calls, emails, and voicemails without having to take their hands off the wheel [14]. An IVN is formed by all of the communications that take place between these sections [10][11].

 The CAN bus protocol establishes a baseline for real-time communication between in-vehicle sensors. The source and destination address for validation are not included in CAN messages because they are communicated from a transmitter to the other units on a bus [12]. As a result, hackers can simply inject any messages into the system, causing it to malfunction. Although for an IVN, the CAN bus proves as a dependable and cost-effective serial bus, during broadcast communication, attackers can quickly get access to the CAN bus [13]. Raising concern is that the CAN bus works without an authentication mechanism, due to which hackers can possibly use an OBD-II port or remote access to the telematics or infotainment systems to manipulate the functionalities of a car [13][14]. For example, hackers can attack the inbuilt Bluetooth, navigation, or sensors of the car resulting in changes to some important functioning of the car. Therefore, in-vehicle CAN network security is critical for the overall security and smooth functioning of modern transportation systems.
 
To counteract threats to vehicle cybersecurity, it is necessary to build countermeasures for numerous attack vectors. One of the ways we may address the problem of cyber-attacks on CAN bus is to use intrusion detection systems [15]. The many attack vectors against cars based on the principles of availability, integrity, and confidentiality of the web world must be thoroughly considered. Intrusion detection systems detect threats by looking for deviations from regular behavior. Intrusion detection systems are prone to false alarms [16]\cite{17}. As a result, businesses must fine-tune IDS products when they initially adopt them. This entails properly setting up their detection mechanism to distinguish between normal network traffic and potentially malicious conduct. A false negative is a far more dangerous IDS mistake in which the IDS overlooks a threat and misidentifies it as regular traffic \cite{17}. In a false negative situation, Security technicians have no indication that a risk is present and do not find out until the networks have been infiltrated in some manner.

This paper presents an intrusion detection approach based on the imbalanced ratios and time-series features, i.e., the frequency of the attacked messages in CAN. The objective of the project is to classify the different attack types and separate them from normal messages based on the CAN data packet. The proposed IDS architecture consists of five stages: data selection, pre-processing, feature selection, model build, and results. In the pre-processing step, we converted and formatted the data type of columns to make it understandable. We selected models from the base model like logistic regression and SVC, then moved to ensemble models like Xgboost and Random Forest and afterward tried the deep learning model such as Feed Forward and LSTM. 

The following points highlight the key contributions of this paper: 
\begin{enumerate}
\item It proposes a CAN-based IDS that can correctly identify the different sorts of cyber-attacks made on intra-vehicle networks.
\item It proposes an attack frequency input feature for the consideration of time-series feature in multi-class classification problems.
\item It proposes a novel anomaly-based IDS based on different AI models starting from base models to ensemble models to deep learning models.
\item It proposes the use of a deep learning model like LSTM, which stores time-series sequence information for multi-class classification problems.
\item It examines and compares the suggested models' performance and overall efficiency on two different input features, one with time-series and other without it.
\item It illustrates that the time-series feature is a key feature consideration for improving model detection accuracy for attacks.
\end{enumerate}

The rest of this document is laid out as follows:
 Section II discusses the Controller area network, CAN messages, and types of CAN attacks. Section III talks about related work in this area. Section IV discusses the insights of the proposed novel IDS. Section V exhibits the architecture and design of the proposed IDS and contains further subsections such as data overview, exploratory data analysis, feature selection, and model selection. Section VI displays and discusses the evaluation and result metrics by comparing the performance of models with and without time-series features. Section VII concludes the paper.

\section{Controller Area Network}
CAN is a ubiquitous vehicle bus standard protocol. Its low cost, flexible design, and highly reliable properties make it useful in the automotive and internet of vehicle world \cite{18}. It works on the broadcasting concept and allows smooth communication between each unit which is connected through CAN. The receiver and sender select the message using the Unique ID property of the message [19].

\subsection{CAN Messages}
A CAN packet frame is a special structure that transmits a sequence of bytes of CAN data over the network. Each transmitted CAN frame has an arbitration identification (ID) field that indicates the packet's priority [19][20]. Packets with a lower ID bit value have a higher priority.

\begin{figure}
     \centering
     \includegraphics[width=8.5cm]{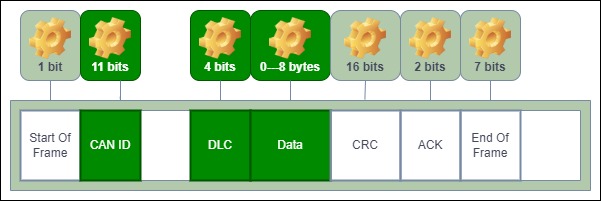}
     \caption{CAN Message format.} \label{packet}
\end{figure}

Of all the fields in CAN, Fig.1 shows the seven important fields of the CAN message.
\begin{itemize}
\item Start Of Frame: The Start of Frame, which is one bit, is being used to sync and alert all nodes that the CAN message transmission is about to begin.
\item CAN ID: A CAN ID which is of 11 bits in the arbitrations field, identifies which ECU the message should be sent to. This parameter determines the priority of the message. In practice, a lower value signifies a bigger priority. The other part of the arbitrations field is the Remote Transmission Request (RTR) of 1bit.
\item DLC: Data Length Code field having 4 bits represents a portion of the control field. 
\item Data: ranging from 0-8 bytes, this field represents the actual data transmission happening to ECUs.
\item CRC: Cyclical Redundancy Check field having 16 bits, represent the code which all receiving units validate before receiving the message.
\item ACK: acknowledge field having 2 bits, is used for getting the flag from the receiver units that they have received the message successfully.
\item EOF: End of Frame field having 7 bits specifies the end of a CAN message.
\end{itemize}	
	
Aforementioned that, CAN packets do not carry any information about the sender or receiver's addresses. As the origin of the packets is not supplied, the recipient nodes are unable to determine if the received packets are meant for them. As a consequence, the recipient ECUs is unable to determine if the packet that it is receiving is valid or not [19][20]. Since the ECUs do not have any authentication or authorization mechanism, resulting in the manipulation of the compromised ECUs by hackers. From the tampered ECU, the attackers have the ability to transmit any unauthorized message to all other units by misusing the CAN bus for message transfer [20]. Koscher et al. [21] explored the wireless intrusions against the automotive system and revealed that it is feasible to hack the CAN bus system and ECUs. They investigated the level of vulnerability of CAN bus and disclosed its sensitivity regarding attacks such as Dos and Spoof. Boyes [22] highlighted the growth of the attack from multiple interfaces and raised various concerns about the security of IVNs. There are lots of other papers calling attention to the security and vulnerability of CAN messages and the importance of having IDS to combat the attacks and to protect the rights of individuals to drive safely and securely. During model training, our project focuses primarily on the highlighted green areas of Figure 1, which are essential components of CAN messages.

\subsection{CAN Attack Types}
Figure 2 shows the known attacks which are covered in this proposed study. In our attack dataset, we explore four main types of known in-vehicle attacks.

\begin{figure}
     \centering
     \includegraphics[width=7cm]{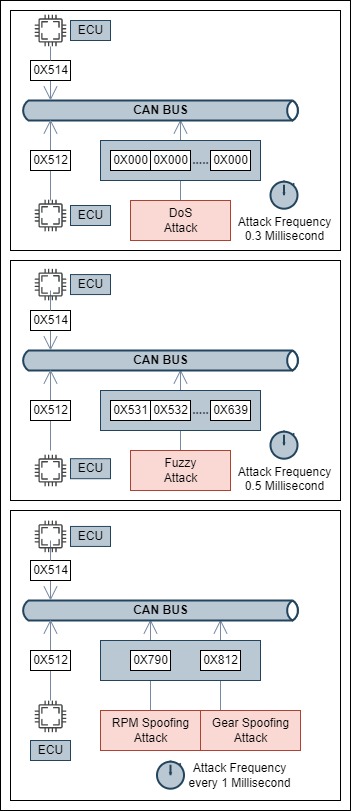}
     \caption{Attack types and frequency of attacks.} \label{packet}
\end{figure}

\begin{itemize}
\item Denial of Service Attack (DoS): DoS attacks entail flooding a host with a massive volume of data in an attempt to overflow it, thereby preventing it from receiving or processing data from authenticated traffic. The Road-Side Unit (RSU) is the most vulnerable section in IVNs. RSUs are an important part of vehicular networks since they identify, monitor, and maintain automobiles and their information \cite{23}.
\item Fuzzy Attack: The technique of introducing random data into software and evaluating the results to uncover possibly exploitable vulnerabilities is what a fuzzy attack is \cite{23}.
\item Spoofing Attack: Spoofing is the act of changing the appearance of a message or identification so that it pretends to come from a reliable, authentic source [10]. Spoofing attacks are further categorized into revolution per minute (RPM) and gear spoofing attack [24].
\end{itemize}	

We have used the attack dataset from the CAR-Hacking dataset provided by Huy Kang Kim et al. [1][25]. The CAN ID identification for DoS attacks with a frequency of 0.3 milliseconds is 0x0000, and a sufficient number of records has been created for Dos Attacks. Random CAN IDs are targeted by a fuzzy attack at a frequency of 0.5 milliseconds. On the two different CAN ID, RPM and Gear Spoofing attacks, records have been developed. The RPM and Gear attack CAN IDs are likely to have the Fuzzy attack as well [25].

\section{Related Work}
Many studies have been proposed on the CAN vulnerability in intra-vehicular networks and the imminent need for IDS since it was understood the necessity of CAN security in light of the rise of the Internet of Vehicles. Huy Kang Kim et al. [1] proposed a novel IDS model for in-vehicle networks, GIDS (GAN-based Intrusion Detection System), using a deep learning model. To improve the speed of their GIDS, they encoded the CAN ID with one-hot encoding for pre-processing. They created the attack data and asserted that their GIDS system covers both known attacks, DoS, Fuzzy, RPM, and Gear spoofing attacks with 100\% accuracy and unknown attacks with 98\% accuracy. Hossain et al. [2] used the LSTM model to classify the different types of attacks such as DoS, Fuzzy, and Spoofing on the CAN-bus in IVNs. They generated the attack data and combined it with normal traffic. They asserted that their IDS is able to detect the attack with 99.995\% accuracy. They tuned the model with different parameters to reach this milestone. To double-check their model's accuracy, they ran it through the Hacking and Countermeasure Research Labs automotive IDS dataset.

Kleberger et al. [3] looked at the security dangers and attacks that can happen in-vehicle network systems and proposed the potential remedies. Qingling Zhao et al. [4] proposed an IDS which detects known (DoS, Fuzzy, RPM, gear) and unknown attacks with different methods considering the combination of classifiers (binary and multi-class), which has been trained on two different networks, ACGAN and OOD. They trained their model on a single dataset having normal, known, and unknown attack category data. They implemented ACGAN and attained 99\% accuracy, while CNN has given the output of around 96\% accuracy. Narayan Khatri et al. [5] proposed a comprehensive blockchain technology usage for Intra-vehicle network security. They insisted on using private blockchain nomenclature for the sensors and ECUs on the inside and the public blockchain method for connecting with the outside network. They highlighted the different types of in-vehicle network protocols such FlexRay, and Automotive Ethernet besides CAN with their properties and security concerns.

Li Yang et al. \cite{tree} \cite{mth} suggested a multi-tiered hybrid intrusion detection system (IDS) that can properly identify the various forms of cyber-attacks made on both internal and external vehicular networks. They combined the two datasets and attempted to identify various threats after combining them. For their IDS training, they employed a variety of tree-structure machine learning models, and got accuracy of around 99.9\% in the Stacking and Xgboost algorithm. In terms of overall performance, their findings reveal that Xgboost was the best method for model training. 

 Omar Y. Al-jarrah et al. [7] provided an organized and thorough examination of state of the art for passenger car intra-vehicle intrusion detection systems (IDSs). They highlighted the earlier surveys which have been done regarding security concerns of the in-vehicle network and the challenges and gaps in in-vehicle networks. They highlighted the pros and cons of different types of available IDS for intra-vehicle networks and argued that no existing single solution which they have covered in their research could provide a complete detection capability of in-vehicle attacks because each detection technique has its own set of flaws and benefits. 
 
 Siti-Farhana Lokman et al. [8] offered an in-depth examination of IDS discovered in the literature, concentrating on the following facets: detection tactics, deployment methodologies, attacking methods, and technological problems. They concluded that adopting a signature and specification-based method is problematic for the CAN protocol because of its structure. However, Learning-based anomaly detection may be a viable detection approach since it can learn from examples and dynamically respond to the CAN environment independent of protocol, vehicle model, or other characteristics.

\section{Proposed Intrusion Detection System}
An intrusion detection system (IDS) is a network traffic monitoring system that detects suspicious behavior and sends out alerts when it is found. There are four categories of IDS.

\begin{itemize}
\item Network Intrusion Detection Systems (NIDS): It is placed inside the network and tracks all the inbound and outbound traffic in the network \cite{26}. 
\item Host Intrusion Detection Systems (HIDS): It is deployed as a host on all computers and devices in the network. The best thing about this IDS is that it can detect attacks from inside the organizations also.
\item Signature-based Intrusion Detection Systems (SIDS):  It looks for a signature for verifying the attacks and compares the signature with an existing database having numerous signature records of attacks [27].
\item Anomaly-based Intrusion Detection Systems (AIDS): It continuously monitors traffic considering the ideal value of bandwidth placed in between external ports and CAN bus and notifies the security team regarding the attacks [27].
\end{itemize}

This research focuses on the identification of attacks by AIDS using a variety of machine learning algorithms. When malicious conduct or anomalous traffic is discovered, AIDS will compile an attack report and send out an attack alert to security team by notifying them. The infrastructure team or administrator can investigate the issue based on alert and can take immediate action to rectify the attack.

Figure 3 shows the architecture of the Intra-vehicle network with various attack interfaces and the location of IDS in the CAN bus. As shown in Fig. 3, All ECUs communicate with each other through the CAN Bus, and different telematics and infotainment devices, including cellular, Bluetooth, music players, and navigation systems, are also connected to the CAN bus and transmit and receive signals [28]. Because CAN lacks authentication, an attacker can engage directly with both OBD Port II and infotainment and telematics devices. Once an attacker sends a message over these devices and ports, they may cause any ECU to malfunction, as seen in the picture, where the red ECU represents the compromised one [28].

\begin{figure}
     \centering
     \includegraphics[width=8cm]{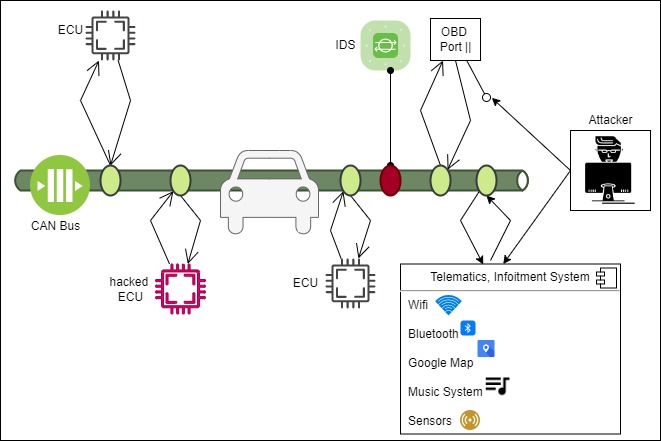}
     \caption{Architecture of Intra-Vehicle Network.} \label{packet}
\end{figure}

IDS is installed between open external interfaces and the CAN bus. All communication from the interfaces will pass from the IDS layer only; if a problem or anomaly is discovered, an alert or report will be issued based on the attack category and severity \cite{29}. Although this paper is confined to alert detection mechanisms, there are certain types of IDS that take action against the source and block the source IP or take other necessary steps in response to the attack.

Machine learning algorithms can play a vital role in detecting attacks. Supervised and ensemble algorithms such as Xgboost and Random Forest as well as deep learning algorithms, aid in categorizing malicious data packets based on known attack patterns. Moreover, Unsupervised learning models can assist in identifying novel malware patterns, as well as detecting and distinguishing them from regular packets \cite{30}.

\subsection{Architecture Description}

\begin{figure}
     \centering
     \includegraphics[width=8cm]{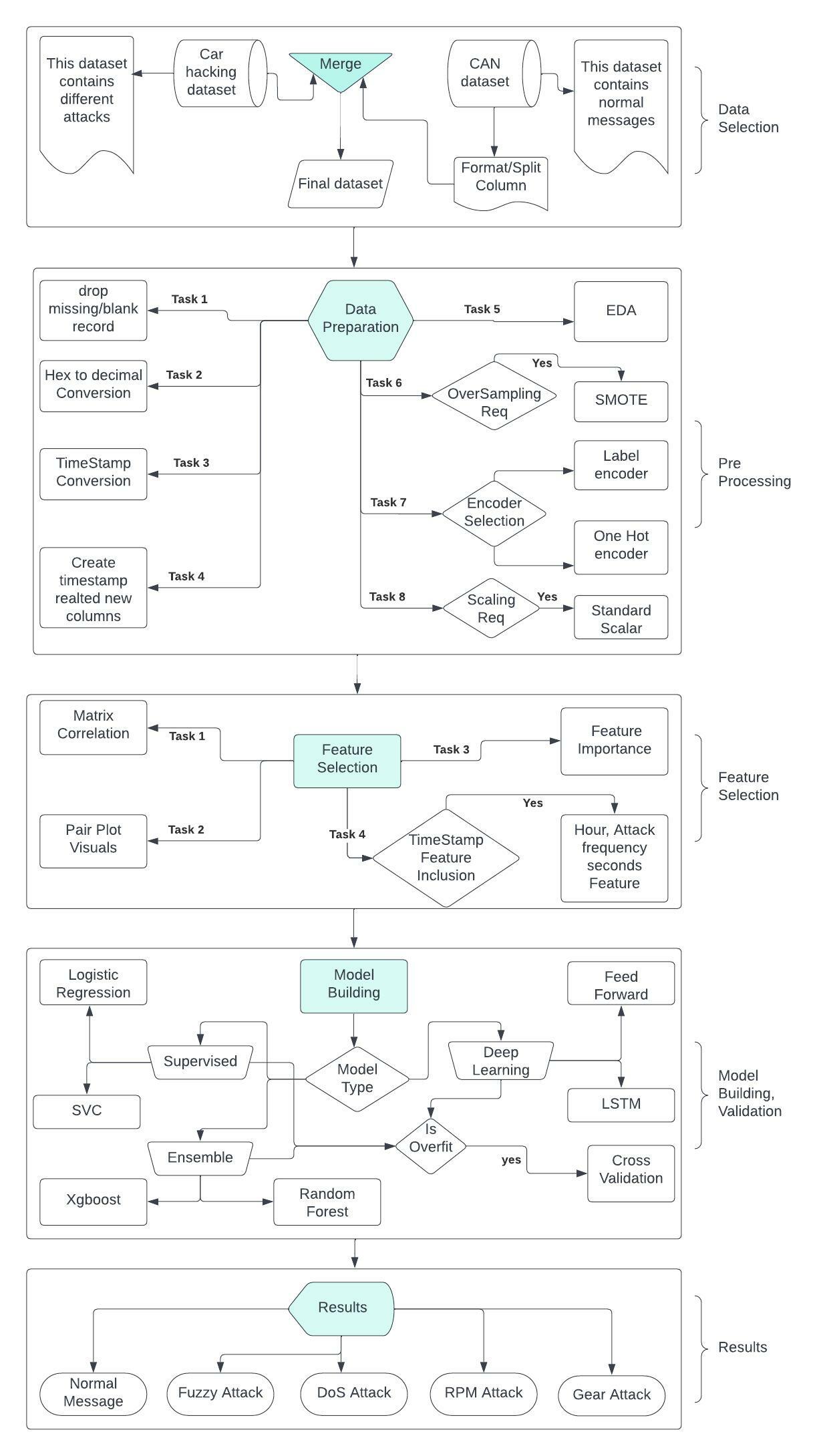}
     \caption{Architecture of the implemented IDS.} \label{packet}
\end{figure}

Figure 4 shows the overall steps which have been taken to design the machine learning-based AIDS. The first process is dataset selection, which entails combining two independent datasets, including the CAN dataset containing normal messages and the attack dataset, to generate a final dataset. The CAN dataset was structured prior to merging it with the attack dataset for the final dataset. Step 2 displays data preparation, which consists of typically eight separate jobs. The first duty is to check for missing/null records, while Task 2 is to convert hexadecimal to decimal, and Task 3 is to convert timestamps to have the DateTime column., Task 4 discusses the establishment of new features based on timestamp columns such as attack-frequency-seconds and hour. The EDA is represented by Task 5. Task 6 discusses the use of oversampling techniques such as SMOTE if necessary. Task 7 depicts categorical value encoding options, depending on the model, we can choose LabelEncoder or OneHotEncoder. In the form of StandardScalar, Task 8 symbolizes scaling. The third stage considers attack-frequency-seconds and hour columns, as well as feature selection and importance. Step 4 shows that many AI models are used to tackle this problem, including supervised (Logistic regression, SVC), ensemble (Xgboost, Random Forest), and deep learning (Feed Forward, LSTM). It also indicates that cross-validation was employed to avoid the issue of overfitting. Step 5 goes over the results and categorizes the attacks as Normal, Fuzzy, DoS, RPM, and Gear attacks.

\subsection{Dataset Overview}
The dataset for this study comprises two different datasets, one with attack category and another with normal messages. The attack dataset files are taken from the CAR-Hacking dataset [25]. This dataset file has four different types of attacks: DoS, Fuzzy, RPM, and Gear spoofing attacks. Normal messages are present in another dataset that contains confidential real-time traffic data provided by the National Research Council Canada. Since the format of the normal message dataset is different from the CAR-Hacking dataset attack files, conversion was required before merging these two files to form a single dataset. The final dataset contains around 350k records where normal records are present in the majority, followed by DoS attack, RPM spoofing, Fuzzy attack, and Gear spoofing attack. Below is the detailed description of both datasets.

Anomaly dataset structure and properties:
\begin{enumerate}

\item Generated by Huy Kang Kim et al. [1] [25] for IDS training.
\item Different attack dataset has been generated based on Attack frequency seconds.
\item Timestamp: the date and time of the recordings.
\item CAN ID: a HEX identification for a CAN message, for example, 02c0
\item DLC: data byte count, ranging from 0 to 8.
\item Data: actual data ranging from 0-7 bytes
\item Flag: T or R, where T denotes an attacked message and R denotes a regular message.
\item Contains data for November-3, 2016.
\end{enumerate}

Normal message dataset structure and properties:
\begin{enumerate}
\item Real-time CAN message traffic dataset provided by the National Research Council Canada (NRC)
\item Contains CAN interface as a column showing two different ports exposed for CAN message.
\end{enumerate}

Contains data for July-19, 2021, some of the records show OBD II messages.

\subsection{Data Preparation}
In all, data preparation entails eight steps. Following the merger of the two datasets, we immediately check for null and empty records and remove them. We are converting the data fields from hexadecimal to decimal format for better assimilation. To acquire the basic Datetime format of “Y-m-d H: M: S.f”, which is more comprehensible for the time series column, we are converting the timestamp feature independently. Then, depending on the CAN ID, we create the attack-frequency-seconds and hour columns to determine the difference between attacked records in the dataset. We do exploratory data analysis when we have sufficient features.
\subsection{Exploratory Data Analysis}
Figure 5 demonstrates that normal messages have the largest distribution, with 200k records, followed by dos and other attacks, which account for 9\% of total records. Both attack and regular messages are recorded between 4 p.m. and 9 p.m., as shown in Fig. 6. It also shows the time intervals for various attacks and typical messages. RPM spoofing attacks occur between 4 p.m. and 5 p.m., gear spoofing attacks occur between 5 p.m. and 6 p.m., Fuzzy attacks occur between 5 p.m. and 7 p.m., DoS attacks occur between 6 p.m. and 8 p.m., and normal messages occur between 7 p.m. and 9 p.m. So, we can conclude that most of the attacks are happening between 4 p.m. to 8 p.m., this is the critical information for our model training.

\begin{figure}
     \centering
     \includegraphics[width=8cm]{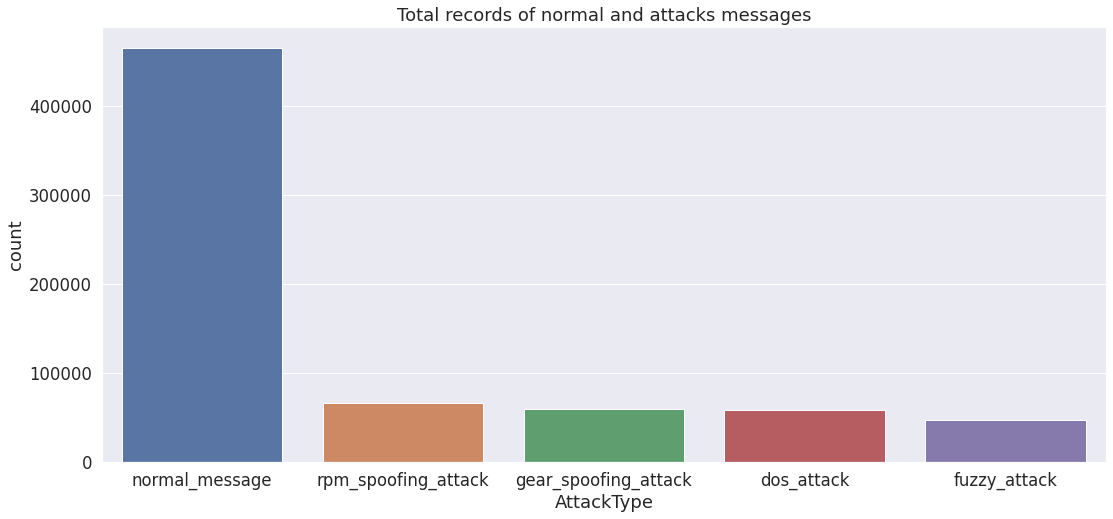}
     \caption{Records distribution of various categories.} \label{packet}
\end{figure}

\begin{figure}
     \centering
     \includegraphics[width=8cm]{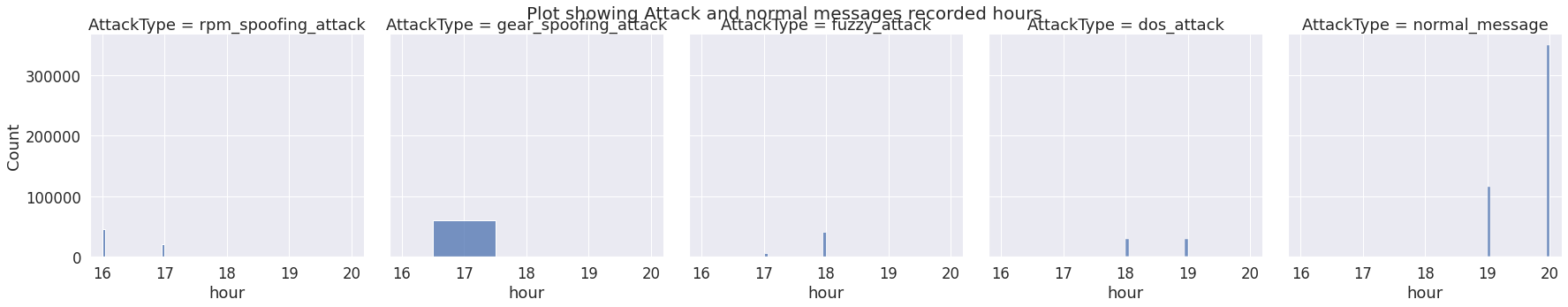}
     \caption{Attack hours of the used dataset.} \label{packet}
\end{figure}

After doing EDA, we were able to determine the ratio of distinct attacks and regular message records, as well as their distribution within the dataset. If necessary, we used the Synthetic Minority Oversampling Technique to do oversampling (SMOTE) [31]. Because we were working on a multinomial classification issue with five output units, it was required to construct an encoder, and we used two separate encoders. The LabelEncoder for supervised and ensemble machine learning models and  OneHotEncoder for neural networks. Later, we used StandardScalar to normalize the data, which is one of the recommended pre-processing steps in machine learning [32].
\subsection{Feature Selection}
To understand the data and model in a better way, we are checking the importance of features by various techniques.

Feature importance scores may be used to aid in data interpretation, but they could be used to rank and pick the most valuable features for a forecasting model [32][33]. Statistical score, which shows a linear relationship between different correlation coefficients for variables, has been used using the Matrix correlation method. From the visual point of view, we used the pair plot to determine the model features importance. However, Feature importance after training the model is something which we are relying on. For supervised-based models like Logistic Regression and SVC, we have retrieved the coeff\_ property, and for tree-based models, our reliable algorithm for the feature importance was Random Forest. Based on the result, we are modifying our independent feature for better training results.

We initially ignore the time-series features, training the model without it as an independent variable; however, we later include the Timestamp feature and retrain the model to verify the impact of time-series features like attack-frequency-seconds and hour on model accuracy. Because both the merged datasets were gathered in separate years, it would have been rather straightforward for the model to categorize the attacks using the day, month, year, and from the feature importance graph as shown in Fig. 7, we deciphered that minute, seconds, and milliseconds properties are not having much importance and do not constitute in model learning, while year, month, the day is not relevant for the model training data is relevant that’s why showing high importance. The hour and attack-frequency-seconds (at\_freq\_sec) features are what we considered to be timestamp features, as illustrated in Fig. 4.

\begin{figure}
     \centering
     \includegraphics[width=7cm]{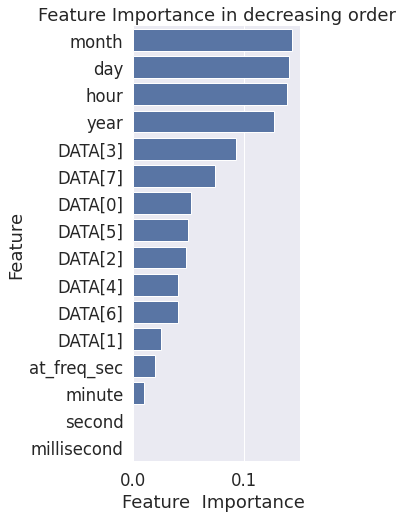}
     \caption{The feature importance generated by Random Forest.} \label{packet}
\end{figure}

The most crucial element from the perspective of feature importance is the timestamp in the form of the hour followed by attack-frequency-seconds (at\_freq\_sec), as illustrated in Fig. 7. Then data field with the column name DATA[3] is important. The DATA[6] column is the one with the least relevant characteristic.

\subsection{Model Selection and Build}
We have used six different types of models for this anomaly problem. 

\subsubsection{Logistic Regression}
Logistic Regression is one of the basic Machine Learning algorithms used to solve classification issues. It is a predictive analytic approach that is based on the probability notion [33]. Generally, it works for the binary classifier, but it also works for the multinomial classification, which is relevant to our problem. The logistic sigmoid function translates the result of logistic regression into a probability (p-value) [33]. We have used the multi-class property as multinomial since our problem is multinomial. Because our problem is multinomial, we have labeled the multi-class attribute as multinomial. In addition, the AUC-ROC curve was plotted using the One-vs-Rest (OVR) method. Regularization L2 has also been applied to guarantee that the mode is not overfitting.

\subsubsection{Support Vector Classification (SVC)}
SVC is also one of the basic supervised machine learning models that work for multinomial classification issues. Its goal is to find the optimal boundary (called a hyperplane) between different classes. SVM, in its most basic version, makes complex data alterations depending on the kernel function supplied, with the purpose of increasing the separation distance between data points. Support Vectors are the data points with the shortest distance to the hyperplane \cite{34}. Since our problem is multi-class, we are using One-vs-One (OVO) approach for plotting the ROC-AUC curve. Regularization has also been implemented to avoid overfitting.

\subsubsection{Random Forest (RF)}
Ensemble Models: Ensemble learning is a broad conceptual approach in machine learning that combines predictions from different models to improve overall predictive performance \cite{35}. Implemented algorithms in this study from ensemble learning are Random Forest and Xgboost.

RF is an ensemble supervised learning model. It has the ability to be utilized for both classification and regression. The trees make up a forest. A forest is thought to be stronger the more trees it has. Random forests generate decision trees using randomly selected data samples, obtain predictions from each tree, then decide on the best option. It also serves as a strong indicator of the value of the feature [36]. The sublime importance of RF is that it barely faces the overfit problem reason behind taking averages of all the tress and decisions. This is one of the strongest reasons we have chosen this algorithm for our research. There are four main steps in the random forest working process:
\begin{itemize}
\item Choose a random selection from a set of data.
\item Create a decision tree for each sample and use it to generate a prediction result.
\item Make a vote for each expected outcome.
\item As the final forecast, choose the prediction with the most votes [36].
\end{itemize}

\subsubsection{Extreme gradient boosting (Xgboost)}
Xgboost is a distributed gradient boosting library that has been tuned for efficiency, flexibility, and portability. It uses the Gradient Boosting framework to create Machine Learning algorithms. Boosting is an ensemble learning strategy for creating a strong classifier from a sequence of poor ones [37]. Xgboost uses parallel tree boosting to tackle a variety of data science issues quickly and accurately [37]. Gradient boosting, Stochastic gradient boosting, and Regularized gradient boosting of both L1 and L2 are a feature of this model \cite{38}.

\subsubsection{Feed Forward (FF)}
Deep Learning model: Deep learning is a subset of machine learning that is fundamentally a neural network of a minimum three-layer [39]. In this work, we have implemented Feed Forward (FF) and Long Short Term Memory (LSTM) models.

A FF neural network is the simplest artificial neural network in which the data flows only in the forward direction, and connections between the perceptron do not create a cycle. The data comes in through the input nodes, passes through the hidden layers, and finally leaves through the output unit. There are no linkages in the networks that would allow information departing the output node to be transferred back into it. Feedforward neural networks are sometimes referred to as a multi-layered set of neurons since all information travels forward only. [40]

\subsubsection{Long Short Term Memory (LSTM)}
The LSTM is a more advanced form of the Recurrent Neural Network (RNN), capable of learning and recalling longer sequences of input data. They are built to cope with data that consists of long sequences of data with 100-to-200-time steps. Because it is one of the best models for our issue, we must train each model with and without timestamp features. Since LSTM considers the timestamp and remembers past sequences, it is predicted to show a significant difference in accuracy when trained with and without timestamp information [41]. Three gates in an LSTM are used to control the flow of information. The forget gate determines which information must be forgotten and which must be passed. The input gate changes the state of the cell by transferring relevant data to the next stage. The next hidden state is determined by the output gate. The previous cell state plus the inputs are used to determine the cell state in LSTM [41].

We decided to start with the most basic algorithm, which is why we picked Logistic regression with multinomial multi-class. We used regularization and varied the penalty value C from 0.1 to 1.0. Later, we chose SVC, which works on determining the best border between classes and enables multi-classes in the (OVO) method. We penalized the model by changing C from 0.1 to 1.0. Then we moved on to ensemble models, with Random Forest being our first choice because it supports multi-class inherently and has the ability to consider both continuous and categorical input features, making it ideal for our situation. Its ability to show the importance and weightage of features also helped us decide on our feature selection process. Then we shifted to a boosting technique, Xgboost, which operates on the principle of stochastic gradient descent with lightning speed and performance. Because our problem was multi-class, we developed it with a multi-softmax objective, and it also offers a wide range of hyperparameter tweaking options \cite{42} \cite{43}. Then, to check the performance of NN for our problem, we moved to deep learning and chose the base model Feedforward Neural Networks (FFNN). Then, for our challenge, we moved to Recurrent Neural Network (RNN) and picked LSTM. This was the best option for distinguishing between model performance with and without time-series features since we wanted to highlight the influence of time-series features on model performance.

\section{Result and Evaluation Metrics}
In machine learning, evaluation of performance is critical. We have used the detection accuracy, F1-score, and confusion matrix score, as evaluation metrics to measure the performance of our model. We have also used the Area Under the Curve (AUC) – Receiver Operating characteristics (ROC) curve evaluation metrics to visualize our models' performance.

\subsection{F1-Score}
The F1-score is a harmonic mean of accuracy and recall, with the best value being 1 and the lowest being 0. Precision and recall both provide an equal proportion to the F1 score. It is calculated as the average of the F1-score of each class in the multi-class situation [44]. It is calculated as:
\begin{equation}
F 1=\frac{2 \times (precision \times recall)}{precision + recall} 
\end{equation}

where precision is defined as the proportion of properly identified positive samples (TP) to the total number of accurately or erroneously classified positive samples. Its value varies from 0.0 lowest to 1.0 highest. It is calculated as:
\begin{equation}
precision=\frac{T P}{T P+F P} 
\end{equation}

Recall is defined as the proportion of properly identified positive samples (TP) to the total number of positive samples which could have been predicted. Its value varies from 0.0 lowest to 1.0 highest. It is calculated as
\begin{equation}
recall=\frac{T P}{T P+F N} 
\end{equation}

Where, TP is true-positive, FP is false-positive, and FN is false-negative. Since our study is on multinomial classification with imbalanced data, if we are not using sampling techniques, the preferred matrix for that reason is F1-score.

As shown in Figs. 8 and 9, we got very good F1-scores for all the implemented models highest being 99\% when we are considering timestamp features, while without those feature considerations, it is limited to 98\%.

\begin{table}[]
\centering%
\caption{F1-SCORE FOR DIFFERENT MODELS 
WITH AND WITHOUT TIME-SERIES FEATURE
}
\setlength\extrarowheight{1pt}
\scalebox{0.85}{\begin{tabular}{|>{\centering\arraybackslash}m{7em}|>{\centering\arraybackslash}m{7em}|>{\centering\arraybackslash}m{7em}|>{\centering\arraybackslash}m{7em}|}
\hline
\textbf{Model Type}            & \textbf{Model Name} & \textbf{F1-Score Without Time Feature} & \textbf{F1-Score With Time Feature} \\ \hline
\multirow{2}{*}{Supervised}    & Logistic Regression & 0.856                                  & 0.991                               \\ \cline{2-4} 
                               & SVC                 & 0.968                                  & 1.00                                \\ \hline
\multirow{2}{*}{Ensemble}      & Random Forest       & 0.972                                  & 0.992                               \\ \cline{2-4} 
                               & Xgboost             & 0.966                                  & 1.00                                \\ \hline
\multirow{2}{*}{Deep Learning} & Feed Forward        & 0.962                                  & 0.986                               \\ \cline{2-4} 
                               & LSTM                & 0.842                                  & 1.00                                \\ \hline
\end{tabular}
}
\label{test}
\end{table}

\begin{figure}
     \centering
     \includegraphics[width=7cm]{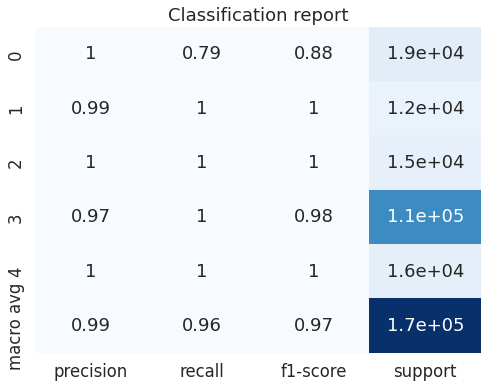}
     \caption{CLASSIFICATION REPORT WITHOUT TIME-SERIES FEATURE.} \label{packet}
\end{figure}

\begin{figure}
     \centering
     \includegraphics[width=7cm]{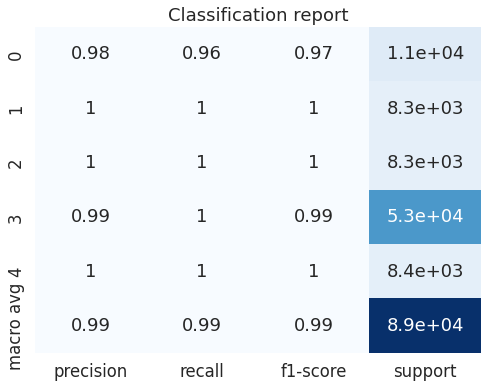}
     \caption{CLASSIFICATION REPORT WITH TIME-SERIES FEATURE.} \label{packet}
\end{figure}

\subsection{AUC-ROC}
The Area Under the Curve (AUC) - ROC curve is a performance evaluation for classification tasks at various threshold levels. AUC indicates the degree or measure of separability, whereas ROC is a probability curve. It indicates how well the model can differentiate between classes [45].

We plot the True Positive Rate (TPR) versus the False Positive Rate (FPR) in our AUC-ROC curve, with FPR on the x-axis and TPR on the y-axis using the One-vs-Rest (OvR) approach.

Formula for TPR, FPR is:
\begin{equation}
T P R=\frac{T P}{T P+F N} 
\end{equation}

\begin{equation}
F P R=\frac{F P}{F P+T N} 
\end{equation}

As shown in Figs. 10 and 11, AUC-ROC values near 1.0 show the models' robustness and superior performance, whereas values near 0.0 show the models' underperformance and weakness [46]. Based on the number of classes available, we are creating AUC-ROC curves for our multi-class classification issue that fit our research. For most of the curves, we recoded values near to 100\%, when considering timestamp features. On the other hand, when we are not using it, we get an AUC-ROC score somewhere around 95\%.

\begin{table}[]
\centering%
\caption{AUC SCORE FOR DIFFERENT MODELS 
WITH AND WITHOUT TIME-SERIES FEATURE
}
\setlength\extrarowheight{1pt}
\scalebox{0.85}{\begin{tabular}{|>{\centering\arraybackslash}m{7em}|>{\centering\arraybackslash}m{7em}|>{\centering\arraybackslash}m{7em}|>{\centering\arraybackslash}m{7em}|}
\hline
\textbf{Model Type}            & \textbf{Model Name} & \textbf{AUC Score Without Time Feature} & \textbf{AUC Score With Time Feature} \\ \hline
\multirow{2}{*}{Supervised}    & Logistic Regression & 0.977                                   & 0.999                                \\ \cline{2-4} 
                               & SVC                 & 0.995                                   & 1.00                                 \\ \hline
\multirow{2}{*}{Ensemble}      & Random Forest       & 0.996                                   & 0.998                                \\ \cline{2-4} 
                               & Xgboost             & 0.996                                   & 1.00                                 \\ \hline
\multirow{2}{*}{Deep Learning} & Feed Forward        & 0.974                                   & 0.986                                \\ \cline{2-4} 
                               & LSTM                & 0.787                                   & 1.00                                 \\ \hline
\end{tabular}
}
\label{test}
\end{table}

\begin{figure}
     \centering
     \includegraphics[width=7cm]{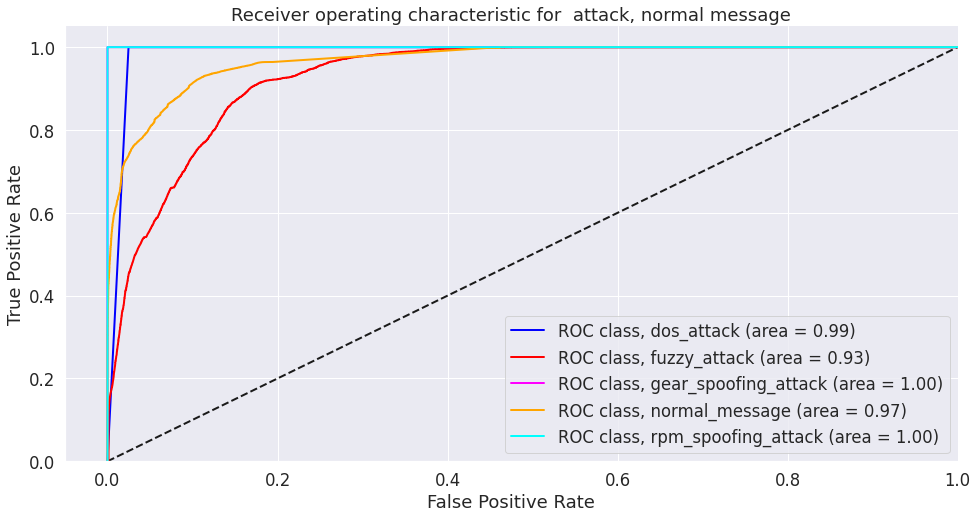}
     \caption{ROC-AUC WITHOUT TIME-SERIES FEATURE.} \label{packet}
\end{figure}

\begin{figure}
     \centering
     \includegraphics[width=7cm]{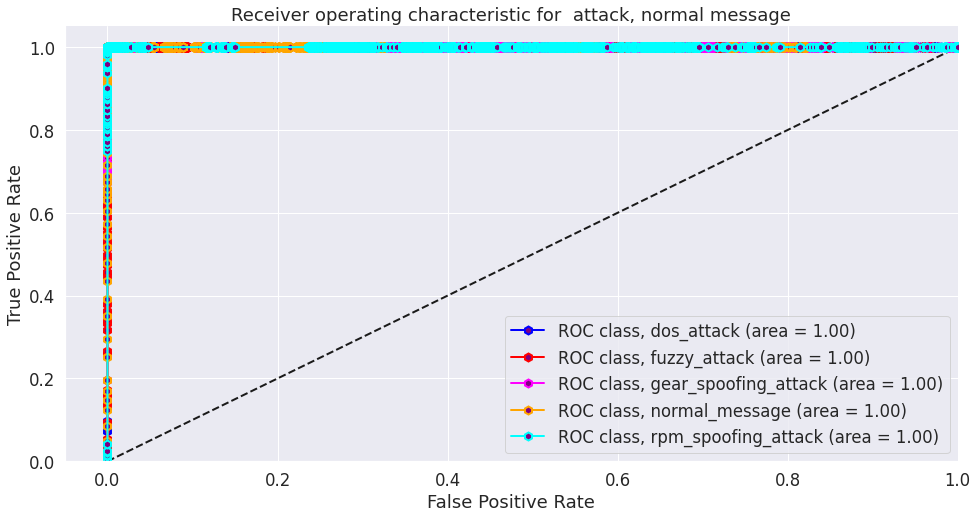}
     \caption{ROC-AUC WITH TIME-SERIES FEATURE.} \label{packet}
\end{figure}

\subsection{Cross-Validation}
Cross-validation is a resampling approach for evaluating machine learning models on a smaller set of data in order to analyze the model and determine whether it is overfitting. It is used to assess the performance of a machine learning model on validation data from training data to generalize the models' performance. The validation dataset will be used to assess the models' performance [45]. When we compare validation loss and accuracy to training loss and accuracy, we can see if our model is underfitting or overfitting [46]. If the model performs well during validation, there is a good probability it will do even better with an unseen test dataset. We have used RepeatedStratifiedKFold and K=3 for validating our models' performance.

\begin{table}[]
\centering%
\caption{MEAN OF CROSS-VALIDATION SCORE FOR DIFFERENT MODELS 
WITH AND WITHOUT TIME-SERIES FEATURE
}
\setlength\extrarowheight{1pt}
\scalebox{0.85}{\begin{tabular}{|>{\centering\arraybackslash}m{7em}|>{\centering\arraybackslash}m{7em}|>{\centering\arraybackslash}m{7em}|>{\centering\arraybackslash}m{7em}|}
\hline
\textbf{Model Type}            & \textbf{Model Name} & \textbf{Cross-Validation Score Without Time Feature} & \textbf{Cross-Validation Score With Time Feature} \\ \hline
\multirow{2}{*}{Supervised}    & Logistic Regression & 0.915                                   & 0.993                                \\ \cline{2-4} 
                               & SVC                 & 0.978                                   & 1.00                                 \\ \hline
\multirow{2}{*}{Ensemble}      & Random Forest       & 0.977                                   & 0.992                                \\ \cline{2-4} 
                               & Xgboost             & 0.974                                   & 1.00                                 \\ \hline
\multirow{2}{*}{Deep Learning} & Feed Forward        & 0.972                                   & 0.994                                \\ \cline{2-4} 
                               & LSTM                & 0.773                                   & 1.00                                 \\ \hline
\end{tabular}
}
\label{test}
\end{table}
\subsection{Confusion Matrix}

The confusion matrix is a performance measurement metric, which shows the overall picture of the prediction results on the classification problem. Since the classes are listed in the rows in the same order as they are in the columns, the correctly categorized items, which are TP, TN, are found on the main diagonal from top left to bottom right, while False values FP, FN are found in the other diagonal from top right to bottom left [47] [48]. In our study, almost all classes are showing True positive rates of 100\% with time-series features, while without them, the True positive rate is 95\%, as shown in Figs. 12 and 13.

\begin{figure}
     \centering
     \includegraphics[width=7cm]{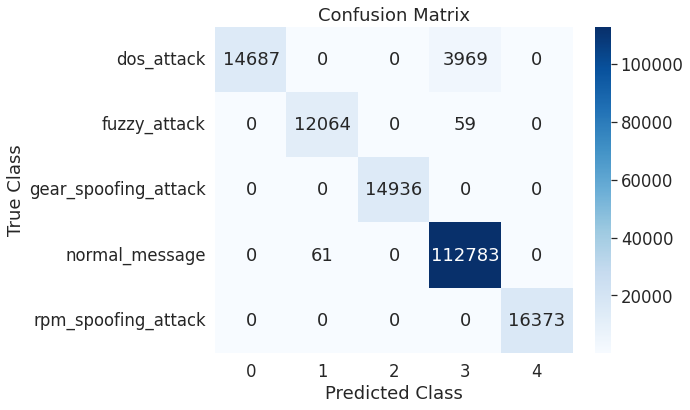}
     \caption{CONFUSION MATRIX WITHOUT TIME-SERIES FEATURE.} \label{packet}
\end{figure}

\begin{figure}
     \centering
     \includegraphics[width=7cm]{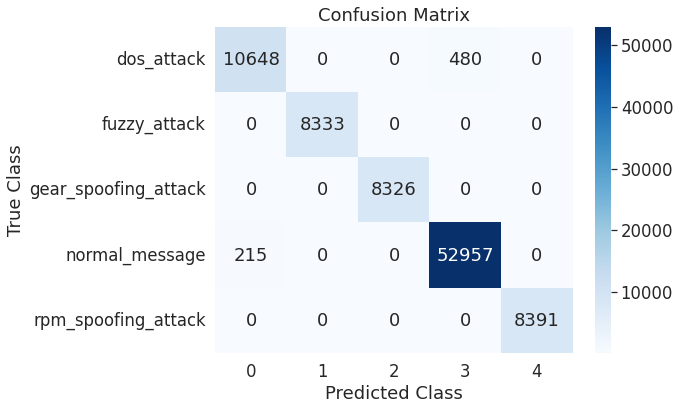}
     \caption{CONFUSION MATRIX WITH TIME-SERIES FEATURE.} \label{packet}
\end{figure}

\subsection{Matrix view of Model Accuracy with and without Time-series}
We used the accuracy score to display it in the matrix below and compare model performance in both scenarios when the model uses timestamp features and when it does not. We can see from the results that when all models are trying to fit data points with timestamp features, which are very important in our dataset, their performance improves.

The best method in our study to highlight the relevance of time-series was the LSTM model, which works well with time-series data. With time-series, it produced 100\% accuracy, but without it, there was a considerable decline in accuracy, and the model halted at 78\% accuracy. Xgboost and SVC, which were deployed, also produced considerable results and proved to be one of the best models with 100 percent accuracy. Furthermore, both of these algorithms work well without the use of time-series data, with an accuracy of about 97 percent. Feed-Forward neural network has justified its worth by giving an accuracy of 99\% with time-series and 98\% without it. Random Forest is one of the methods that has provided the greatest accuracy for non-time-series training, with a 98\% accuracy. With 99\% accuracy, it performed admirably for the time-series feature.

Thus LSTM, Xgboost, and SVC proved as the best model to develop IDS with 100\% accuracy and F1-score to detect the attacks on the CAN. Furthermore, accuracy, F1-score have been improved significantly by considering time-series features which justifies the importance of time-series information for IDS.

\begin{table}[]
\centering%
\caption{PERFORMANCE ACCURACY FOR DIFFERENT MODELS 
WITH AND WITHOUT TIME-SERIES FEATURE
}
\setlength\extrarowheight{1pt}
\scalebox{0.85}{\begin{tabular}{|>{\centering\arraybackslash}m{7em}|>{\centering\arraybackslash}m{7em}|>{\centering\arraybackslash}m{7em}|>{\centering\arraybackslash}m{7em}|}
\hline
\textbf{Model Type}            & \textbf{Model Name} & \textbf{Accuracy -  Without Time Feature (\%)} & \textbf{Accuracy - With Time Feature (\%)} \\ \hline
\multirow{2}{*}{Supervised}    & Logistic Regression & 92.60                                   & 99.15                                \\ \cline{2-4} 
                               & SVC                 & 97.20                                   & 100                                  \\ \hline
\multirow{2}{*}{Ensemble}      & Random Forest       & 98.20                                   & 99.32                                \\ \cline{2-4} 
                               & Xgboost             & 97.40                                   & 100                                  \\ \hline
\multirow{2}{*}{Deep Learning} & Feed Forward        & 97.12                                   & 99.45                                \\ \cline{2-4} 
                               & LSTM                & 78.25                                   & 100                                  \\ \hline
\end{tabular}
}
\label{test}
\end{table}
\section*{Conclusion}
In this paper, we propose an AIDS based on Machine learning models to detect the anomaly on the CAN-bus in intra-vehicular networks. An In-depth discussion about the CAN bus vulnerability and the reason why AIDS is needed in this area has been covered in our study. The known attack category and its types of which are possible in the CAN-bus and its consequences on the vehicle and driver were discussed. We propose the framework and design of IDS which is capable of detecting the anomalies with 99\% accuracy. Since there were two datasets from distinct sources, the attack dataset was generated in a different year than the normal messages real traffic dataset, which was gathered in a different year after merging the dataset; thus, our goal was to identify the relevant features from both datasets. Our problem is a Multi-class Classification problem having timestamp series data. The major goal of this paper is to demonstrate the value of timestamp characteristics in model training for detecting and verifying the four different types of attacks. We utilized attack-frequency-seconds and the hour as one of the timestamp features. As such, forming a correlation between these two datasets of two different years' data was difficult. We used various machine learning models, ranging from supervised to ensemble to deep learning, to compare the outcomes with and without timestamp features. The importance of the time-series feature is evident in all of the evaluation metrics results and assessment measures. 

\section*{Acknowledgment}
I would like to acknowledge and thank Prof. Abdallah Shami and Mr.
Li Yang from OC2 Lab, Department of Electrical and Computer Engineering,
Western University, for their support and help throughout this work.

\ifCLASSOPTIONcaptionsoff
  \newpage
\fi



\end{document}